\documentclass[prb,twocolumn,showpacs,preprintnumbers,amsmath,amssymb,a4paper]{revtex4}

\usepackage{graphicx}
\usepackage{dcolumn}
\usepackage{bm}

\begin{document}



\title{Equation of State, Stability, Anisotropy and Nonlinear Elasticity of Diamond-Cubic (ZB) Silicon by Phonon Imaging at High Pressure}

\author{F. Decremps$~^{a}$,\cite{email}, L. Belliard$~^{b}$, M. Gauthier$~^{a}$ and B. Perrin$~^{b}$}

\affiliation{$~^{a}$Institut de Min\'{e}ralogie et Physique des
Milieux Condens\'{e}s, Universit\'{e} Pierre et Marie Curie-Paris 6,
CNRS UMR 7590, 140 rue de Lourmel, 75015 Paris, France.\\
$~^{b}$Institut des NanoSciences de Paris,
Universit\'{e} Pierre et Marie Curie-Paris 6, CNRS UMR 7588, 140 rue
de Lourmel, 75015 Paris, France.}

\date{\today}

\begin{abstract}

Experimental phonon imaging in diamond anvils cell is demonstrated to be an adequate tool to extract the complete set of elastic constants of single-crystalline silicon up to the ZB$\rightarrow \beta-$Sn transition (10 GPa). Contrary to what was commonly admitted, we demonstrate that the development of the strain-energy density in terms of strains cannot be stopped, for silicon, after the terms containing the third-order elastic constants. Nonlinear elasticity, degree of anisotropy and pressure-induced mechanical stability of the cubic silicon structure are thus revisited and investigated in more detail.

\end{abstract}


\maketitle

\section{Introduction}

The determination of interatomic forces in solids with respect to atomic displacements is one of the most gentle and direct way to probe the short-range repulsive potential. Experimental study of elastic properties under high pressure has thus become an essential part of solid-state-physics, although such measurements are still a nontrivial task. For example, the measurement of elastic constants as a function of the pressure (i.e. the crystal volume) can serve to derived a precise and hypothesis-free equation of state (p,V,T). It is also useful to understand structural and electronic instabilities in solids. Last but not least, the well-known accuracy of ultrasonics measurements has been shown to allow the determination of both second- and higher-order elastic constants, giving rise to the characterization of nonlinear and anharmonic properties of solids.

Already studied for more than fifty years through the pressure variation of sound waves under moderated pressures (less than 1 GPa), the determination of nonlinear properties has recently attracted great interest, in conjonction with the development of nanoelectromechanical and nanoelectronic devices. In such systems, the large surface contribution to the total energy modifies the properties and, in particular, raises the intrinsic nonlinear properties. In a recent paper, \L{}opuszy\ifmmode \acute{n}\else \'{n}\fi{}ski \emph{et al}\cite{lopus} have shown through DFT calculations that the knowledge of pressure derivatives of elastic constants is a clear prerequisite to a complete explanation of semiconductors nanostructures properties. Tang \emph{et al}\cite{tang} came to the same conclusion through a calculation of lattice dynamics in silicon nanostructure, one of the most interesting system for applications as sensors or communication technologies.

Extrapolating the pioneering work of McSkimin and Andreatch in 1964 on the elastic properties of cubic silicon\cite{mcskimin1} and germanium\cite{mcskimin2} up to 0.2 GPa, the hydrostatic pressure dependence of the 3 independent Second-Order Elastic Constants (SOEC) of covalent semiconductors has always been considered to be linear. Rigorously, these experimental results only give an access to the determination of few Third-Order Elastic Constants C$_{IJK}$ (TOEC) combinations, constants which can simply be expressible in terms of SOEC pressure derivatives dC$_{IJ}$/dp. To determine the six independent TOEC of a cubic crystal, one also required experiments under uniaxial stress. However, difficulties inherent to the non-hydrostatic technique\cite{stress} are known to produce doubtful conclusions\cite{bogardus} and/or severe disagreements between authors. Even more conflicting are the inconsistencies between hydrostatic and uniaxial data in a same work\cite{johal}. To avoid any controversies, we here restrict our attention to the nonlinear effects of a single crystal of silicon under hydrostatic pressures.

Reaching higher pressure as possible is crucial since the rather subtle contribution of high-order elastic constants is known to increase with increasing pressure. From an experimental point of view, our main goal was thus to improve our knowledge on the elasticity of cubic silicon through an highly accurate sonar measurement over the largest pressure domain of observation. To follow this objective, we have used the purpose-built technique of picosecond acoustics phonon imaging at high pressure. This method, described in the first part of the article, combines the great accuracy of ultrasonic laser technique with the capabilities of diamond anvils cell (DAC). It allows the measurement of the complete set of elastic constants up to the transition pressure (10 GPa) of a defect-free and undoped single-crystalline silicon sample\cite{hall}. In a second part, we show that such measurements may serve as a starting point to discuss the validity of the linear regression hypothesis of C$_{IJ}$(p), commonly admitted. Nonlinearity and anisotropy of the Si elastic properties as a function of pressure are also discussed in terms of phase stability. This may be used as a stringent test of accuracy (or even validity) for state-of-the-art simulations, or even as a significant information to improve the modeling of silicon-based nanoelectromechanical and nanoelectronic devices.

\section{Phonon imaging in DAC : arguments and principle}

The main reason why acoustic measurements on single crystal at high pressure are quite sparse in literature is technical. Traditionally, the frequency of sound waves generated by ultrasonic transducers lies typically in the MHz range, so that any sonar-based measurements would require millimetric sample sizes. Such high dimensions render the piezoelectric transducer technique to only be performed when combined with a large-volume cell~\cite{baosheng}. This, consequently, limits the pressure domain. To overcome this limitation, an attempt to implement acoustic GHz interferometry~\cite{spetzler} with DAC has been done. This promising method is however still limited in pressure due to the presence of diffraction effect or acoustic phase shifts at the diamond-sample interface under non-hydrostatic stresses~\cite{anderson}. The third experimental scheme is the well-known Brillouin scattering, an optical method that can easily be adapted to DAC~\cite{polian}, but which is unsuitable in the case of silicon : in Brillouin scattering method, transparent sample can exclusively be studied.

More recently, picosecond acoustics\cite{antonelli, perrin, tatiana} has been developed to enable highly accurate mechanical properties of opaque materials as metal liquids~\cite{decremps1}, polycrystals~\cite{decremps2}, or single-crystal~\cite{decremps3} in DAC. In this method, the measurement is similar to the pulse-echo ultrasonic technique (travel time determination) with the advantage of the optical methods (no contact ; no bonding effect). The principle of picosecond acoustics in DAC, described in references~\cite{decremps1, decremps3}, is based on the absorption of the light pump pulse which sets up a local thermal stress near the surface. In the acoustic far field diffraction limit (i.e. the laser illuminates the sample with a spot size much higher than the film thickness), only longitudinal acoustic strain field is created. In order to circumvent the "shearless" problem, we further focus our attention on the potentiality of picosecond ultrasonics to generate and detect shear waves in sample embedded in a DAC. Keynote is to generate lateral compressive stresses (producing internal diffraction) through a minimization of the source area with respect to the characteristic acoustic wavelength.

Original development of acoustics using small source and point detector have revealed some striking properties of phonons in anisotropic materials, such as phonon-focusing effect~\cite{maris1}. In crystals, the direction of acoustic energy flow does generally not coincide with the wave vector direction $\textbf{k}$. Consequently, a source producing an uniform angular distribution of $\textbf{k}$ will generate an anisotropic propagation of elastic energy. Considering a crystal with plane parallel surfaces, a point acoustic source at one surface will produce non-uniform focusing pattern at the opposite face. This produces patterns related to the complex topology of the group velocity surface (also called wave surface). Such intriguing imaging can however be calculated from elastic theory (within the continuum model) taking into account the relation between directions of $\textbf{k}$ (or phase velocity $\textbf{v}=\omega(\textbf{k})/\textbf{k})$ and directions of energy flow (or group velocity $\textbf{v}^e=\partial \omega(\textbf{k}) / \partial \textbf{k}$). Experimentally, the earlier images of flux energy pattern, obtained in heat-pulse experiments~\cite{maris2}, have shown that a massive amount of data may be extracted from successively recording acoustic wavefront snapshots at different times. In recent years, the technique of picosecond laser acoustics has been shown to be well suited for such purpose : femtosecond duration of the laser pulse and the possibility to focalize the beams onto the surface plane of the sample give rise to an outstanding resolution in space and time. In particular, the measurements of group velocities using picosecond acoustics has been shown to allow the determination of the complete set of elastic constants of anisotropic sample at ambient pressure and high temperature~\cite{audoin1}. Following this approach, and taking into account the considerable progress made the last five years in ultrafast acoustics~\cite{rossignol}, we have here implemented the wavefront imaging method in the case of picosecond acoustics in DAC.

\section{Experimental set-up}

Ultrashort pulses of 100 fs are generated every 12.6 ns by a Ti:Sapphire laser (800 nm). The laser beam is split into pump and probe beams.
The pump is focused on one surface of the sample, whereas the probe is focused on the opposite one. As soon as the pump laser pulse reaches the surface, it creates a sudden and small
temperature rise (of about 1 K). The corresponding thermal stress generated by thermal expansion relaxes by launching acoustic strain fields.
After propagation along the sample, both thermal and acoustic effects alter the optical reflectivity of the sample in two ways: the photo-elastic effect,
and the surface displacement (as the acoustic echo reaches the surface). The first modification contributes to the change of both real and imaginary parts of the reflectance, whereas the
second one only modifies the imaginary contribution. In a pure thermo-elastic model, the time and space reflectivity change $\Delta r(t)$ can be represented as a function of the photo-elastic coefficient  $\partial n / \partial z$ and the surface displacement $u_0(t)$ as~:

\begin{equation}
{\frac{\Delta r(t)}{r_0} = ik_0 \{2u_0(t) + \frac{\partial n}{\partial z} \frac{4n}{1-n^2} \int_{0}^{+\infty} \eta(z,t) e^{2ik_0nz}dz} \}
\label{eq01}
\end{equation}

where n is the optical index of the sample.

The variation of reflectivity as a function of time is detected through the
intensity modification of the probe delayed from the pump with a different optical path length. The uncertainty on the absolute position of the delay line being less than 1 $\mu$m, the determination of the pump/probe time delay is better than 20 fs. The detection is carried out by a
stabilized Michelson interferometer which allows the determination of the reflectivity imaginary part change~\cite{perrin1}.
Microscope objectives are here necessary in order to focus both pump and probe beams down to 3 $\mu$m (objectives with a typical working distance of about 20 mm and an optical aperture of 0.42, well adapted to the DAC environment, have been used). Finally, an image is obtained by scanning the reflectivity of the whole sample surface through the displacement of the probe objective in the plane perpendicular to the beam using a X-Y piezoelectric stage (giving rise to images of 100 $\mu$m x 100 $\mu$m, with an accuracy on the absolute position better than 1 $\mu$m). We emphasize that, albeit the experimental set-up is similar as the one used in a sonar configuration (see Ref~\cite{decremps3}), the surface displacement of the sample is here determined by scanning the surface at fixed pump-probe delay (i.e. at a given time along the acoustic propagation).

Extracted from a large single-crystal of silicon, a thin platelet oriented along [100] with surface of about 70*70 $\mu$m$^2$ was mechanically polished. A thin film (50 nm) of Al was sputtered on both sides to serve as transducers. This sample was loaded into the experimental volume of the DAC. Neon was used as pressure transmitting medium. Using the well known longitudinal velocity of Si along the [100] direction at ambient conditions~\cite{mcskimin1}, the thickness of the crystal platelet was measured to be 42.2(1) $\mu$m from picosecond measurement in the classical configuration (i. e. scanning the delay at fixed pump-probe relative position). Pressure was classically measured using the fluorescence emission of a 5 $\mu$m ruby sphere~\cite{mao} placed close to the sample in the gasket hole. The accuracy was better than 0.1 GPa at the maximum pressure reached.

\section{Results and discussion}

\subsection{Elasticity at ambient conditions}

\begin{figure}[h]
\includegraphics[width=8. cm]{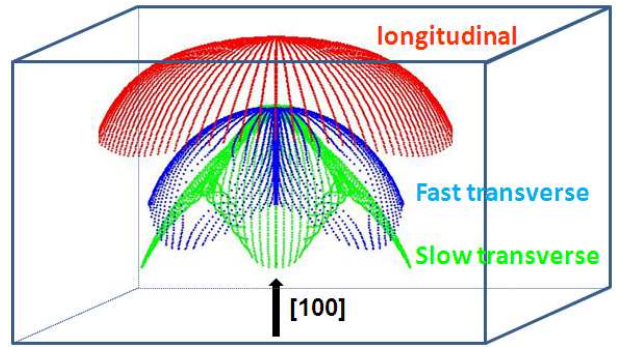}
\caption{\label{fig01a} Calculated group velocity surfaces near the [100] symetry axis of ZB-cubic silicon. Red, blue and green dashed lines correspond respectively to the longitudinal, fast and slow transversal group velocities.}
\end{figure}

Using the density $\rho$=2.331 g.cm$^{-3}$ and a set of initial elastic constants values, the simulated slices of the wave surface in the (100) plane for silicon at ambient conditions are obtained in two steps. Firstly, the Christoffel equation is solved for a set of wave vectors $\textbf{k}$ lying within 45 degrees (cubic symmetry) around the [100] crystallographic direction. The slowness curves are then generated for the three acoustical polarizations and, in a second step, used to calculate from ray theory~\cite{auld} the wave front curves within a surface cut in the plane (100). In Figure~\ref{fig01a}, a typical 3-D phonon imaging pattern for the three acoustic polarizations is given.

\begin{figure}[h]
\includegraphics[width=8. cm]{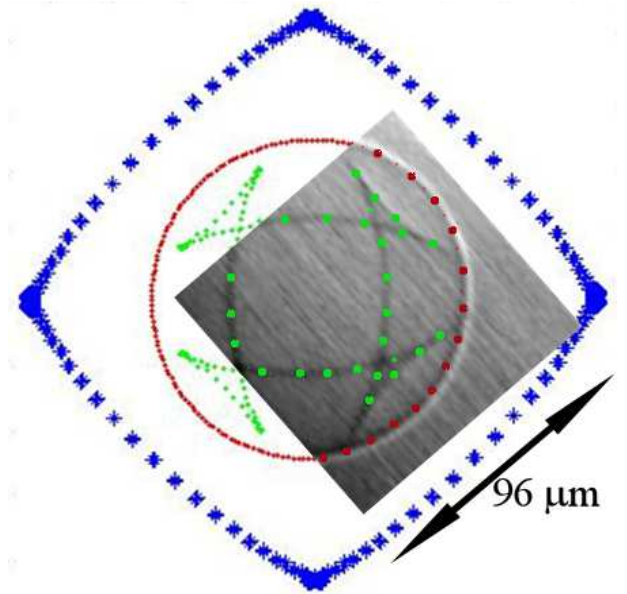}
\caption{\label{fig01b} Experimental phonon imaging pattern in the (100) plane of silicon at ambient conditions (pump-probe delay time of 0.4 ns). Red, blue and green dashed lines correspond respectively to the longitudinal, fast and slow transversal group velocities using C$_{11}$= 165.7 GPa,  C$_{12}$= 63.9 GPa and C$_{44}$= 79.5 GPa.}
\end{figure}

The experimental and calculated intercepts between 3-D wave front and (100) plane of silicon at ambient conditions is shown in Figure~\ref{fig01b}. The simulated process of the wave surfaces, where the elastic constants are the fitting parameters, well reproduces the experimental pattern and gives rise to an elastic tensor in very good agreement with previously published data~\cite{mcskimin1}. We here emphasize that, for a given thermodynamical condition (here ambient), different patterns observed for different pump-probe delays can be used to renew the fitting process in order to determine the complete set of elastic constants with a higher accuracy.

\begin{figure}[h]
\includegraphics[width=8. cm]{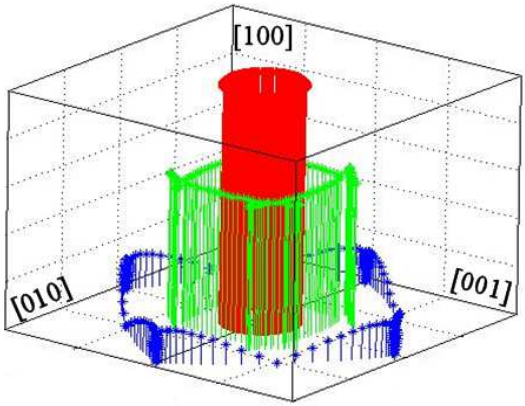}
\caption{\label{fig01c} Calculation of polarization vectors (eigenvectors of the Christoffel equation) projected along the [100] direction, perpendicular to the surface where the probe is focalized. Red, blue and green dashed lines correspond respectively to the longitudinal, fast and slow transversal polarization.}
\end{figure}

Note that the fast transversal mode was not experimentally detectable and thus not been taken into account in the elastic constants fitting process. The absence of this mode (polarized into the (100) plane) can be easily understood using Eq.~\ref{eq01}, where the imaginary part is mainly dominated by the surface displacement perpendicular to the surface. Consequently, using our interferometric configuration, any surface displacement could be detected except when it is perpendicular to the surface. A calculation of a [100] projection of polarization vectors, shown in Figure~\ref{fig01c}, illustrates the absence of the fast transversal mode for which the corresponding displacement is almost zero.

\subsection{Nonlinearity and anisotropy at high pressure}\label{sec1}

\begin{figure}[h]
\includegraphics[width=7.1 cm]{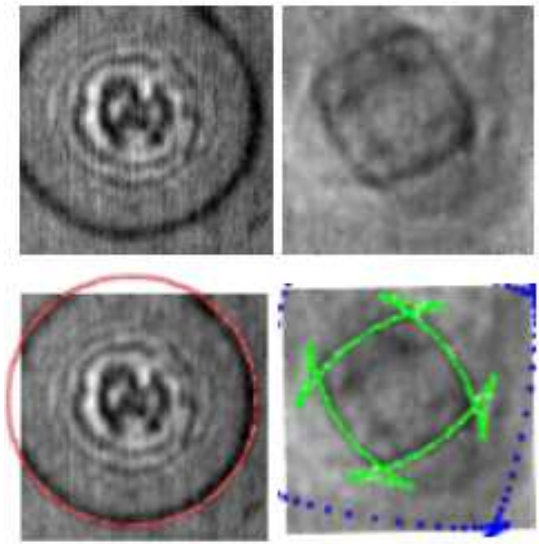}
\caption{\label{fig01d} Top : experimental phonon imaging patterns in the (100) plane of silicon at 7.75 GPa at two different probe-pump delays. Bottom : same as top with superimposed calculation curves for longitudinal, fast and slow transversal group velocities (red, blue and green dashed lines respectively) using C$_{11}$= 196.9 GPa,  C$_{12}$= 104 GPa and C$_{44}$= 80 GPa.}
\end{figure}

At high pressure, a step forward of the previous section, typical experimental and calculated slices of the wave surface in the (100) plane are given in Figure~\ref{fig01d}.

Taking into account the pressure dependence of the density, we have here used a classical procedure to determine the thickness of the crystal plate at each pressure. The basic idea can be briefly described as follows. Assuming that length and density are known at a given pressure p (ambient pressure, for example), all velocities and elastic moduli can be directly deduced from the picosecond measurements. Using p-values of density and thickness, and p+$\Delta$p wavefront snapshots for different pump-probe delay, the length and the density of the sample at p+$\Delta$p can be computed. These values are then used to deduce a  first approximation of the elastic moduli at the new pressure p+$\Delta$p. The values of the latter are finally used as starting points in an iterative process, until convergence is reached. This process is quite robust since the variation of the sound velocity is mainly due to that of the elastic moduli (i.e. its dependence on length and density is only of second order).

\begin{figure}[!h]
\includegraphics[width=8. cm]{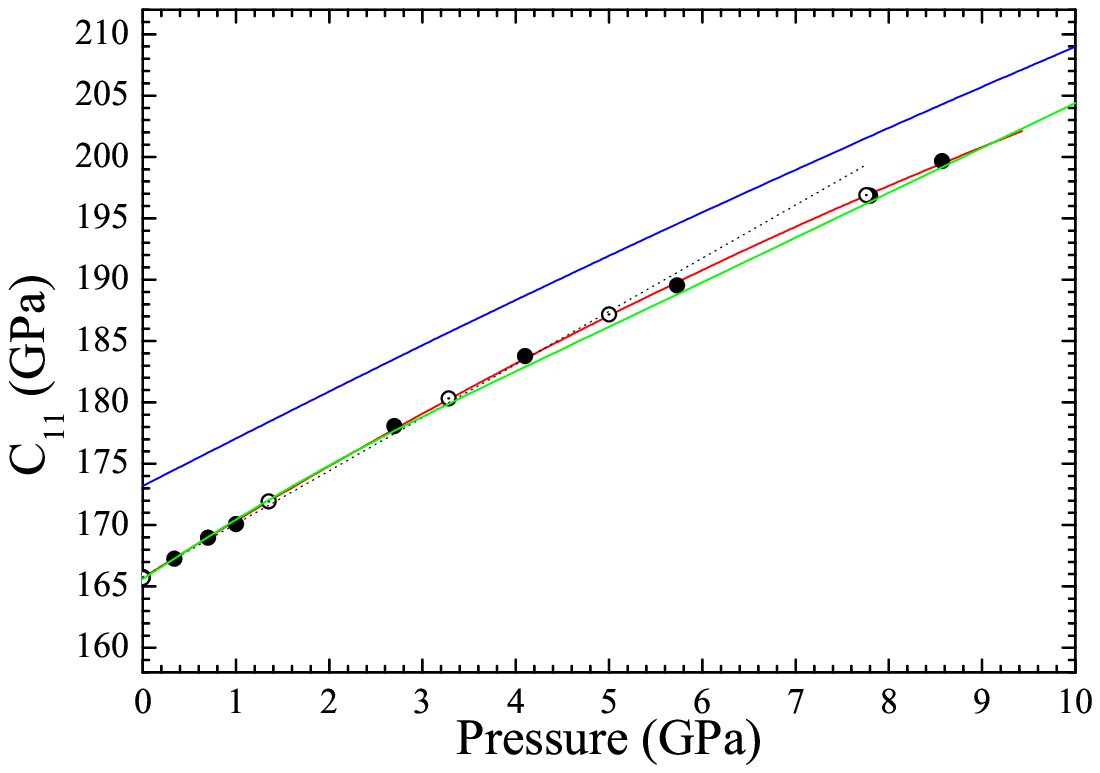}
\includegraphics[width=8. cm]{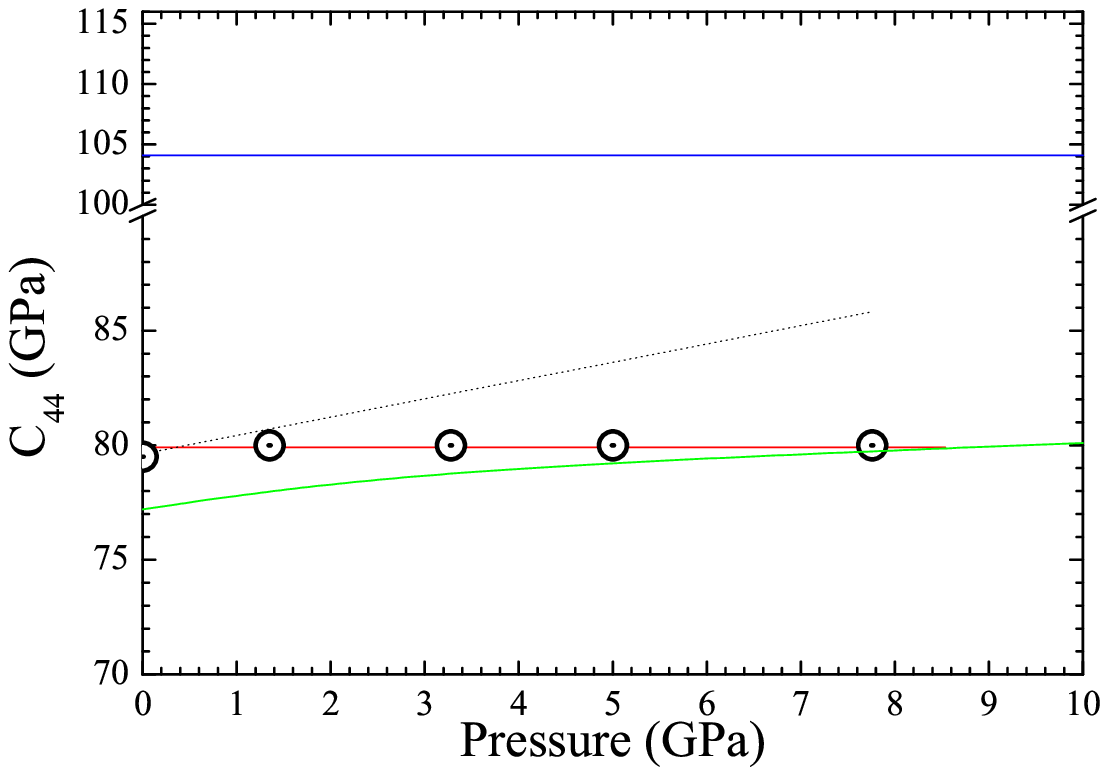}
\includegraphics[width=8. cm]{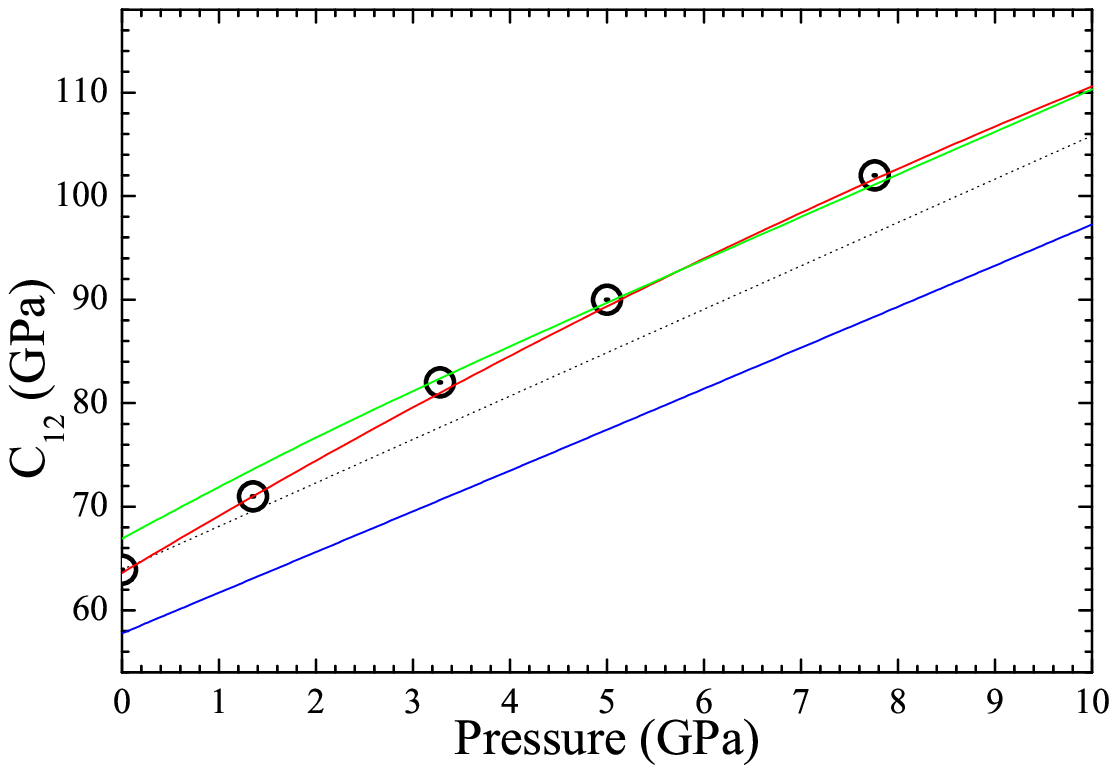}
\caption{\label{fig02} Pressure dependence of cubic silicon elastic constants C$_{11}$ (up), C$_{44}$ (middle) and C$_{12}$ (down) at 300 K. The data for C$_{11}$ corresponds to experiment of phonon imaging (open circles) completed by an experiment in a classical high pressure picosecond set-up\cite{decremps3}. Present results (with uncertainty given by the symbol size) are compared with previous data (blue solid line : Tight-binding DFT calculations\cite{cohen}, green solid line : LDA-DFT calculations\cite{karki}, and dot black line : extrapolation of low-pressure ultrasonics data\cite{mcskimin1}). The red solid line corresponds to a second-order polynomial fitting of the phonon imaging data.}
\end{figure}

Figures~\ref{fig02} show elastic constants of cubic diamond (ZB) silicon as a function of pressure up to the ZB$\rightarrow \beta-$Sn transition (10 GPa). An excellent agreement is observed with computations results using DFT-LDA total-energy method\cite{karki}. The comparison is also very good with tight-binding simulations\cite{cohen}, except for C$_{44}$ which calculated ambient pressure value is far from the well established value of 80 GPa\cite{landolt}. Whereas this discrepancy may be due to an inadequate model parametrization, a good agreement between theoretical and experimental pressure derivative of C$_{44}$ is nevertheless observed. Present data and 1960's ultrasonics results\cite{mcskimin1} are also in excellent agreement (in the pressure domain where the comparison can be done, say less than 0.2 GPa). In this pressure range, the infinitesimal strain theory does not hold anymore and, within the accuracy of the measurements, second-order moduli C$_{IJ}$ are observed to vary linearly with pressure.
However, a comparison between our data at higher pressure (typically more than 3 GPa) and a linear extrapolation of the McSkimmin\cite{mcskimin1} results is worse. Beyond the classical quasiharmonic approximation, the pressure dependence of the third-order moduli C$_{IJK}$ of silicon is here needed to interpret such disagreement. A second-order polynomial regression of present data yields to (data in GPa)~:

C$_{11}$(p)=165.7+4.73p-0.09p$^2$;

C$_{12}$(p)=63.60+5.78p-0.12p$^2$;

C$_{44}$ is however unaffected by a variation of pressure, lying around 80 GPa within the experimental uncertainty.

Consequently, the pressure change of three independent relations for C$_{IJK}$ (here labeled $E_i(p)$) is calculated\cite{birch} to be (data in GPa)~:

E$_1$(p)=C$_{111}$+2C$_{112}$=-1860+28p;

E$_2$(p)=2C$_{112}$+C$_{123}$=-1120+31p;

E$_3$(p)=C$_{144}$+2C$_{166}$=-747.

A good agreement is observed (see Table~\ref{tab01}) between the zero-pressure values E$_i(0)$ with previously published data.

\begin{table}[h]
\caption{\label{tab01} Comparison of experimental (exp) and calculated (calc) relations E$_i$ for C$_{IJK}$ (in GPa) of silicon at ambient conditions.}
\begin{ruledtabular}
\begin{tabular}{cccc}
reference & E$_1(0)$ &  E$_2(0)$  & E$_3(0)$ \\
\hline
Present results  (exp) & -1860 & -1120 &  -747 \\
Reference \cite{mcskimin1} (exp) & -1729 & -997 &  -610 \\
Reference \cite{philip} (exp) & -1854 & -1064 &  -687 \\
Reference \cite{lopus} (calc) & -1600 & -1014 & -580  \\
Reference \cite{nielsen} (calc) & -1710 & -960 & -580  \\
\end{tabular}
\end{ruledtabular}
\end{table}

Whereas an evaluation of any of the six independent C$_{IJK}$ specifically is beyond the capability of elastic measurements under hydrostatic pressure, present results clearly point out the contribution of the fourth-order elastic constants (FOEC) C$_{IJKL}$. In qualitative agreement with our present observation, Prasad \emph{et al}\cite{prasad} have evaluated C$_{1111}$ to lie around 12 TPa, high value which entails the FOEC contribution to be substantial at high pressure. This clearly demonstrates the need to account for high-order elastic constants in the study of the anharmonic properties of silicon, of most importance for simulations (test of validity for explicit crystal potential), or for applied physics (harmonics generation and distortion effects) as sound waves propagating in silicon-based devices.

\begin{figure}
\includegraphics[width=8.5 cm]{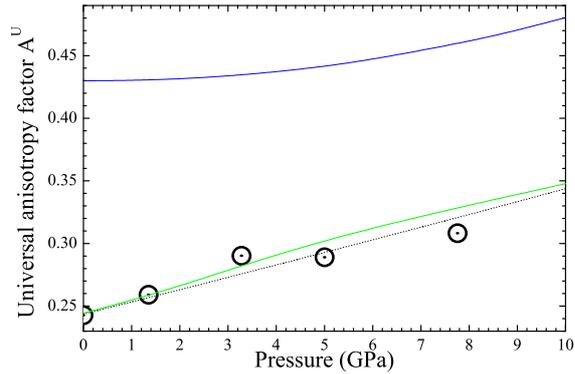}
\caption{\label{fig03} Universal elastic anisotropy factor $A^U$ of ZB silicon as a function of the pressure (open circles) compared with previous data (blue solid line : Tight-binding DFT calculations\cite{cohen}, green solid line : LDA-DFT calculations\cite{karki}, and dot black line : extrapolation of low-pressure ultrasonics data\cite{mcskimin1}).}
\end{figure}

As well as nonlinearity, elastic-anisotropy is also known to shed light on the mechanical properties of solids as dislocation dynamics, plastic deformations or structural stability. For cubic-symmetry as silicon, the 1930's Zener\cite{zener} definition of the acoustical anisotropy $A$=2C$_{44}$/(C$_{11}$-C$_{12}$) has been recently revisited\cite{shiva} in order to quantify the single crystal elastic anisotropy from an universal point of view $A^U=6/5(\sqrt{A}-1/\sqrt{A})^{2}$ (note that for an isotropic crystal, $A$=1 but $A^U$=0). As illustrated in Fig.~\ref{fig03}, the degree of elastic anisotropy of silicon is increasing with increasing pressure. This result well agrees with ab-initio calculations\cite{cohen, karki} and could be related to the observation that the transverse phonon frequency (i.e. the gruneisen parameter) has been found to be the unique mode that decreases with pressure\cite{mizushima}.

\subsection{Equation of state and structural stability}

For cubic crystals, the bulk modulus B is a simple function of the elastic constants~:

B=(C$_{11}$+2C$_{12}$)/3

\begin{figure}
\includegraphics[width=8. cm]{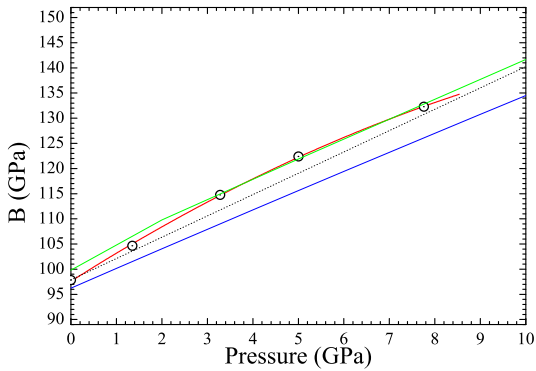}
\caption{\label{fig04} Pressure dependence of bulk moduli in
single-crystalline ZB silicon. The
uncertainties in pressure and bulk modulus are within the symbol
size. Present data are compared with previously published results (blue solid line : Tight-binding DFT calculations\cite{cohen}, green solid line : LDA-DFT calculations\cite{karki}, and dot black line : extrapolation of low-pressure ultrasonics data\cite{mcskimin1}).}
\end{figure}

Using the experimental values of C$_{IJ}$(p), highly accurate and free-hypothesis equation of state B(p) has been determined (see Fig.~\ref{fig04}).

A second-order polynomial curve well reproduces the experimental pressure dependence of the bulk modulus with~:

B(p)=97.60+5.74p-0.16p$^2$

This result in excellent agreement with a fourth-order regression of the elastic energy with respect to the strain giving~:

B'=dB/dp=-(E$_{1}$+2E$_{2}$)/(9B)=5.08-0.24p

Lattice stability and sequence of phase transformations in semiconductors can be understood in the framework of phonons instabilities\cite{ozolin}, which for purely covalent compounds as silicon can be expressed in terms of mechanical stability of the stressed lattice. In a cubic crystal, the condition of a positive density of elastic energy transforms to the following criteria~:

B$_1$=C$_{11}$+2C$_{12}>$0;

B$_2$=C$_{11}$-C$_{12}>$0;

B$_3$=C$_{44}>$0.

B$_1$ is directly connected to the bulk modulus and does not allow to discuss the phase stability in terms of symmetry change from cubic (ZB) to tetragonal ($\beta-$Sn). However,  B$_2$ and B$_3$ are referred to bulk shear and tetragonal shear moduli respectively.

\begin{figure}
\includegraphics[width=8. cm]{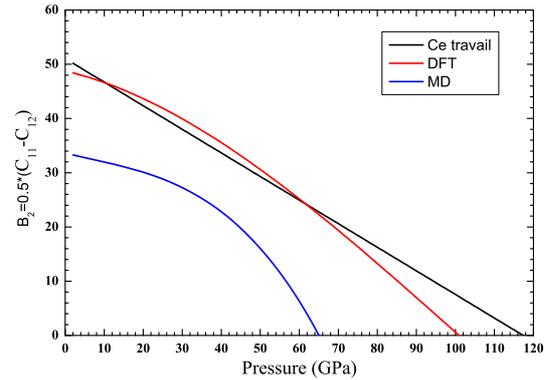}
\caption{\label{fig05} Elastic stability criteria B$_2$ as a function of pressure. Extrapolation of present results (black line) are compared with previously published calculations (blue solid line : molecular dynamics simulations\cite{mizushima} and red solid line : LDA-DFT calculations\cite{karki} .}
\end{figure}

Theoretical studies of ZB structural stability of silicon have been undertaken through DFT\cite{karki} and molecular dynamics simulations\cite{mizushima}. In both works, the determination of the vibrational distortions at high pressure gives a critical pressure (quite higher than the physical transition one) at which the ideal lattice become unstable against homogeneous tetragonal shear deformation, say B$_2$. In our case, a tentative of extrapolation of experimental results can only be done for B$_2$ since the pressure dependence of C$_{44}$ was not enough accurate to be decently interpolate by other than a constant value. As can be seen in Fig.~\ref{fig05}, the violation of B$_2$ occurs at around 120 GPa, in good agrement with theoretical calculations.

This result can be interpreted as the following : silicon, as mainly other tetrahedrally bonded semiconductors\cite{brazkhin}, is mainly stabilized by noncentral covalent interactions. From that point of view, the ZB$\rightarrow \beta-$Sn path transition has intrinsically a martensitic nature, ideal transformation that can not be effectively observed due to kinetic contributions. Consequently, the deformation of ZB-silicon under hydrostatic pressure can been seen to be driven by a tetragonal shear strain, but without reaching the structural instability.

\section{Conclusion}

We have presented a detailed experimental study of elastic constants of cubic ZB silicon as a function of hydrostatic pressure. Phonon imaging in diamond anvils cell, an original experimental purpose-built set-up, allows to determine the subtle effect of high-order elastic constants and demonstrates the failure of the linear regression approximation dC$_{IJ}$/dp=constant. From this study, nonlinear acoustics, anisotropy, equation of state as well as structural stability of a pure single-crystal of  silicon have been extracted with an outstanding accuracy. So far, this technique is likely to have an impact on the study of dynamics of all single-crystals at high density.


This work has been supported by an ANR contract (No. ANR-08-BLAN-0109-01).


\begin{references}
\bibitem{anderson} The travel time measurements is here affected by phase shifts due to the presence of interfaces (diamond anvil) or acoustic coupling. This effect is even more pronounced when the pressure transmitting medium solidifies.
\end{references}
\end{document}